\documentclass[onecolumn,preprintnumbers,amssymb,amsmath,superscriptaddress,letterpaper]{revtex4}[10pt]
\usepackage{graphicx}
\usepackage{dcolumn}
\usepackage{bm}
\usepackage{axodraw}
\usepackage{pstricks}
\usepackage{color}
\begin{document} 
\input epsf.tex

\def\draft{
}

\newcommand{\beq}{\begin{eqnarray}}
\newcommand{\eeq}{\end{eqnarray}}
\newcommand{\nn}{\nonumber}
\def\ltap{\ \raise.3ex\hbox{$<$\kern-.75em\lower1ex\hbox{$\sim$}}\ }
\def\gtap{\ \raise.3ex\hbox{$>$\kern-.75em\lower1ex\hbox{$\sim$}}\ }
\def\CO{{\cal O}}
\def\CL{{\cal L}}
\def\CM{{\cal M}}
\def\tr{{\rm\ Tr}}
\def\CO{{\cal O}} 
\def\CL{{\cal L}}
\def\CN{{\cal N}}
\def\mpl{M_{\rm Pl}}
\newcommand{\bel}[1]{\be\label{#1}}
\def\al{\alpha}
\def\bt{\beta}
\def\eps{\epsilon}
\def\eg{{\it e.g.}}
\def\ie{{\it i.e.}}
\def\mn{{\mu\nu}}
\newcommand{\rep}[1]{{\bf #1}}
\def\be{\begin{equation}}
\def\ee{\end{equation}}
\def\bea{\begin{eqnarray}}
\def\eea{\end{eqnarray}}
\newcommand{\eref}[1]{(\ref{#1})}
\newcommand{\Eref}[1]{Eq.~(\ref{#1})}
\newcommand{\gsim}{ \mathop{}_{\textstyle \sim}^{\textstyle >} }
\newcommand{\lsim}{ \mathop{}_{\textstyle \sim}^{\textstyle <}}
\newcommand{\vev}[1]{ \left\langle {#1} \right\rangle }
\newcommand{\bra}[1]{ \langle {#1} | }
\newcommand{\ket}[1]{ | {#1} \rangle }
\newcommand{\erg}{{\rm erg}}
\newcommand{\eV}{{\rm eV}}
\newcommand{\ev}{{\rm eV}}
\newcommand{\kev}{{\rm keV}}
\newcommand{\Mev}{{\rm MeV}}
\newcommand{\gev}{{\rm GeV}}
\newcommand{\tev}{{\rm TeV}}
\newcommand{\mev}{{\rm MeV}}
\newcommand{\meV}{{\rm meV}}
\newcommand{\mnu}{\ensuremath{m_\nu}}
\newcommand{\nnu}{\ensuremath{n_\nu}}
\newcommand{\mlr}{\ensuremath{m_{lr}}}
\newcommand{\acc}{\ensuremath{{\cal A}}}
\newcommand{\mav}{MaVaNs}
\newcommand{\kt}{\ensuremath{k_{tach}}}
\newcommand{\zt}{\ensuremath{z_{tach}}}
\newcommand{\degree}{^\circ}
\newcommand{\la}{\,\rlap{\raise 0.5ex\hbox{$<$}}{\lower 1.0ex\hbox{$\sim$}}\,}
\newcommand{\ga}{\,\rlap{\raise 0.5ex\hbox{$>$}}{\lower 1.0ex\hbox{$\sim$}}\,}

%
%
\newcommand{\s}{{\rm ~s}}
\newcommand{\sr}{{\rm ~sr}}
\newcommand{\kms}{{\rm ~km/s}}
\newcommand{\cm}{{\rm ~cm}}
\newcommand{\MHz}{{\rm ~MHz}}
\newcommand{\GHz}{{\rm ~GHz}}
\newcommand{\Hz}{{\rm ~Hz}}
\newcommand{\K}{{\rm ~K}}
\newcommand{\mK}{\rm ~mK}
\newcommand{\microK}{\mu{\rm K}}
\newcommand{\pc}{{\rm ~pc}}
\newcommand{\kpc}{{\rm ~kpc}}
\newcommand{\sigmav}{\langle\sigma v\rangle}
\renewcommand{\v}[1]{\mathbf{#1}}
\newcommand{\disc}[1]{{\bf #1}} 
\newcommand{\fluxunit}{{\rm ~ph~cm^{-2}~s^{-1}}}
\def\aj{AJ}                   
\def\apj{ApJ}                 
\def\apjl{ApJ}                
\def\apjs{ApJS}               
\def\aap{A\&A}                
\def\mnras{Mon. Not. Roy. Astron. Soc.}

\title{Exciting Dark Matter and the INTEGRAL/SPI 511 keV signal}
\author{Douglas P. Finkbeiner}
\affiliation{Harvard-Smithsonian Center for Astrophysics, 60 Garden
  St, Cambridge, MA  02138}
\author{Neal Weiner}
\affiliation{Center for Cosmology and Particle Physics,\\
  Department of Physics, New York University,\\
New York, NY 10003, USA}
\preprint{ }
\date{\today}
\begin{abstract}

We propose a WIMP candidate with an ``excited state'' 1-2 MeV above
the ground state, which may be collisionally excited and de-excites by
$e^+e^-$ pair emission.  By converting its kinetic energy into pairs,
such a particle could produce a substantial fraction of the 511 keV
line observed by INTEGRAL/SPI in the inner Milky Way.  Only a small
fraction of the WIMPs have sufficient energy to excite, and that fraction drops
sharply with galactocentric radius, naturally yielding a radial
cutoff, as observed.  Even if the scattering probability in the inner
kpc is $\ll 1\%$ per Hubble time, enough power is available to produce
the $\sim3\times10^{42}$ pairs per second observed in the Galactic
bulge.  We specify the parameters of a pseudo-Dirac fermion designed to
explain the positron signal, and find that it annihilates chiefly to
$e^+e^-$ and freezes out with the correct relic density.  We discuss
possible observational consequences of this model.

\end{abstract}
\maketitle
\twocolumngrid
\section{Introduction}

One of the central unsolved problems of both particle physics and
cosmology is the nature of the dark matter. Although well motivated
candidates exist, for instance a weakly-interacting massive particle
(WIMP) or axion, they have thus far
eluded definitive detection. This makes indirect signals, such as
anomalous particle production in the Galaxy, especially interesting as
a possible indicator of the physics of dark matter.

\subsection{The Positron Annihilation Signal}
In 1970, the first detection of a gamma-ray line in the Galactic
center found a line center at $473\pm30\kev$, causing the authors to
doubt its true origin
\cite{Johnson:1972}. Leventhal \cite{Leventhal:1973} soon pointed out
that the spectrum was consistent with positronium emission, because
positronium continuum could confuse the line fit in a low-resolution
spectrograph, but worried that the implied
annihilation of $7\times10^{42}$ pairs/s was several orders of magnitude
larger than his estimate of the pair creation rate from cosmic ray
interactions with the interstellar medium (ISM).  Since then, other
balloon and space
missions such as \emph{SMM}\cite{Share:1988}, \emph{CGRO}/OSSE
\cite{Kinzer:2001ba} and most recently the SPI spectrograph
\cite{Attie:2003,Vedrenne:2003} aboard the International Gamma-Ray
Astrophysics Laboratory (INTEGRAL) have greatly refined
these measurements \cite{Knodlseder:2003sv,Weidenspointner:2006nu} (see
\cite{Teegarden:2005} for a summary of previous measurements).  We now
know that the line center is $510.954\pm0.075\kev$, consistent with
the unshifted $e^+e^-$ annihilation line \cite{Churazov:2004as} but still
find the estimated annihilation rate of $\sim3\times10^{42}$ $e^+e^-$ pairs
per second \cite{Knodlseder:2003sv,Weidenspointner:2006nu} surprising.
The origin of these positrons remains a mystery.

The OSSE data showed a bulge and a disk component, with $\sim25\%$ of the
flux in the disk\cite{Kinzer:2001ba}.
The most recent INTEGRAL data show mainly the bulge component,
approximated by a Gaussian with FWHM $6\degree$ \cite{Weidenspointner:2007}.
The data are
consistent with a disk component several times fainter than the bulge,
when Galactic CO maps are used as a disk template.  The spatial
distribution of the ortho-Ps (positronium) 3-photon continuum and 511
keV line are mutually consistent, with $92\pm 9\%$ of the pairs
annihilating through Ps \cite{Weidenspointner:2006nu}. The faintness of the
disk component constrains the expected (conventional) $e^+$ production
modes: cascades from cosmic ray interaction with the ISM, and
nucleosynthetic processes during supernovae both would have a disk
component.  In order to achieve the observed bulge/disk ratio with type
Ia SNe, one must assume that the positrons from the disk migrate to the
bulge before they annihilate; otherwise the bulge signal would only be
$\sim10\%$ of that observed \cite{Prantzos:2005pz}.

Indeed, Kalemci et al.\cite{Kalemci:2006bz} find that the
quantity of pairs alone rules out SNe as the only source.  Using SPI
observations of SN 1006 (thought to be type Ia), they derive a
$3\sigma$ upper limit of 7.5\% for the positron escape fraction.
An escape fraction of 12\% would be required to produce all of
the positrons in the Galaxy, but even then they would be distributed
throughout the disk, like the 1.8 MeV line from
$^{26}$Al \cite{Plueschke:2002}, rather than
concentrated in the Galactic center.  Hypernovae
remain a possibility,
if there are more than 0.02 SN 2003dh-like events (SNe type Ic) per century
\cite{Casse:2003fh}.  GRB rates of one per $8\times10^4$ yr $\times
(E_{GRB}/10^{51}\erg)$ are also sufficient \cite{Parizot:2005}.

It is estimated that microquasars could contribute enough positrons to
produce approximately 1/3 of the annihilation signal
\cite{Guessoum:2006}, but again, it is not clear that their spatial
distribution is so strongly biased towards the Galactic bulge. 

In summary, no obvious astrophysical explanation for this positron
production exists; proposals in the literature have
difficulty explaining the size or spatial distribution of the
signal, and often both.  While it is still possible that conventional
(or more exotic) astrophysical sources could produce the entire
signal, the concentration of dark matter in the Galactic center
motivates consideration of an alternative hypothesis: that some
property of dark matter is responsible for the signal.

\subsection{Dark Matter}
In recent years, WIMP annihilation scenarios have been invoked to
explain a number of high energy astrophysical
phenomena (see \cite{Bertone:2004pz} for a review).  
It is possible to create the pairs
directly from WIMP annihilation if the WIMP mass is a few MeV
\cite{Boehm:2003bt}, but this mass range has no theoretical motivation.
In extensions to the Standard Model which addresses the hierarchy problem, such as supersymmetry, 
the WIMP mass is usually above the weak scale (100 GeV - 1 TeV) and the annihilation
products come out at energies very large compared to $m_e$.  In the
absence of a dense environment where a cascade can develop (column
densities of $\sim 10^{27} \cm^{-2}$ are required for the
$\gamma\rightarrow e^+e^-$ step), there is no way to partition the
energy of these particles into the many thousands of pairs required
for each WIMP annihilation.  Consequently, it has generally been
concluded that weak scale dark matter has nothing to do with the
observed positronium annihilation signal. However, we shall see that
WIMPs in this mass range could still play a role: they could convert
WIMP kinetic energy into pairs via scattering.

The kinetic energy of a 500 GeV WIMP moving at 500 km/s is $> 511$
keV.  If the WIMP has something
analogous to an excited state -- for whatever reason -- inelastic
scattering could occur, raising
one or both of the WIMPs to this excited state.  If the energy splitting is
more than $2 m_e$, the decay back to ground state would likely
be accompanied by the emission of an $e^+e^-$ pair.  One tantalizing
feature of this mechanism is that it provides access to the vast
reservoir of kinetic energy in the WIMPs (roughly $10^{60}$ erg for
the Milky Way), such
that bleeding off $3 \times 10^{42}~\Mev/\rm{s} = 5\times 10^{36}$ erg/s
reduces the kinetic energy by a negligible amount, even over the age
of the Universe.  Indeed, the unexplained excess in the bulge is
much smaller than that.  Therefore, this scattering would have negligible
kinematic effect on the dark matter halo, except perhaps in its innermost
parts. 

In this paper, we propose that the positrons in the center of the
Galaxy arise precisely from this process of ``exciting'' dark matter
(XDM). Remarkably, such a phenomenon is quite natural from the
perspective of particle physics in theories with approximate symmetries in the dark sector, which are generally accompanied by nearly degenerate states, which can serve as excited states. Although the MSSM dark matter
candidate, the neutralino, does not exhibit such a property, this is
simply the result of the restricted field content of the MSSM, with no
additional approximate symmetries present. Other WIMPs, both arising
in supersymmetry and from strong dynamics easily give this
phenomenology -- and usually without upsetting the desirable properties
of WIMP models.

The layout of this paper is as follows: in section \ref{sec:pheno} we
discuss the basic properties of such a scenario, and constraints on
the model from the annihilation signal. In section \ref{sec:model}, we
review a simple example from particle physics with the required
properties, utilizing a single pseudo-Dirac fermion as dark matter. In
this scenario we will also comment on the annihilation rate into
standard model particles and other direct and indirect tests of the
model.

\section{Constraints from the 511 keV signal}
While the spectral line and continuum data from SPI are robust when
averaged over the entire inner Galaxy, the spectrum in each spatial
pixel is noisy.  In order to stabilize the fit and robustly determine
the total positronium (Ps) annihilation rate in the bulge, the SPI maps are
cross-correlated with spatial templates.  Because the Ps emission in
the inner Galaxy does not follow the stellar mass distribution, the
choice of templates is somewhat arbitrary.  Representing the disk with
a CO map, and the bulge with a Gaussian, Weidenspointner et
al. \cite{Weidenspointner:2006nu} obtain continuum Ps fluxes in their ``analysis bands'' of
$f_b=0.86^{+0.15}_{-0.13}\times10^{-3}\fluxunit$
and
$f_{CO}= 1.92^{+0.49}_{-0.48}\times10^{-3}\fluxunit$
for the bulge and disk, respectively, finding 31\% of the emission is
in the bulge.  Because the bulge component covers less solid angle
than the disk, its surface brightness is much higher.  Using the full
SPI response matrix they then obtain a total Ps continuum (bulge plus
disk) of
$(3.11 \pm 0.56)\times 10^{-3}\fluxunit$
and line flux of
$(9.35\pm0.54)\times10^{-4}\fluxunit$.
For a Galactocentric distance of 8 kpc = $2.4\times10^{22} \cm$, this implies a total pair annihilation rate of $1.1\times10^{43}\s^{-1}$,
or $3.4\times10^{42}\s^{-1}$ for the bulge.  Because of order 10\% of
that is explained by SNe \cite{Prantzos:2005pz}, we take
$3\times10^{42}\s^{-1}$ as the excess pair creation rate in the
Galactic bulge to be explained.

SPI also measures the width of the annihilation line to be
$2.37\pm0.25~\kev$ FWHM, which is consistent with a warm ($7000 < T_e
< 4\times10^4\K$) weakly ionized
medium \cite{Churazov:2004as}.  Knowledge of the state of the ISM in the
inner Galaxy may be used to constrain the injection energy of the $e^+$
to less than a few MeV \cite{Jean:2005af}, which agrees with the
constraints set by Beacom et al. using internal bremsstrahlung
\cite{Beacom:2004pe} and propagation \cite{Beacom:2005qv}.  In our
model, the few MeV constraint applies to the splitting between the states, not the mass of the WIMPs themselves. 

\label{sec:pheno}

\subsection{Form of the cross section}
In this section we use elementary considerations of the phase-space
density and expected $v$-dependence of the cross section to derive
constraints on the ratio of the splitting to the mass $(\delta/M)$
from the slope of the 511 keV emission.

The velocity-averaged rate coefficient (i.e. number of scatterings per
time per density) $\sigmav$ is 
\be
\sigmav(\v{r}) = \int d^3v d^3v' f(\v{v,r}) f(\v{v',r})
   \sigma(v_{rel})v_{rel}
\ee
where $f(\v{v,r})$ is the phase-space density of particles at position
$\v{r}$ and velocity $\v{v}$, $\sigma(v_{rel})$ is the inelastic
scattering cross section as
a function of the velocity difference $v_{rel}=|\v{v}-\v{v'}|$,
which vanishes below threshold ($v_{rel} < v_{thresh}$), and the
integrals are taken over all velocities.

In order to make an estimate of the parameters involved, we
approximate the velocity distribution as isotropic with no
particle-particle velocity correlations (caused by e.g. net rotation
of the DM halo).  This reduces $\sigmav$ and $f$ to functions of
scalar $r$ and $v$.
The cross section of the proposed pseudo-Dirac fermion has a weak
dependence on velocity, so we first focus on a simplified form of
$\sigmav$ where the velocity dependence is entirely encoded in the
phase space of the final state particles, i.e., where the form of the
cross section is
\bel{eqn:sigmavsimple}
\sigma v_{rel} = 
\begin{cases}
\sigma_{mr} \sqrt{v_{rel}^2 - 4\delta/M_\chi} & v_{rel}^2 > 4\delta/M_\chi \\
                      0                       & v_{rel}^2 \le 4\delta/M_\chi
\end{cases}
\ee
Here $\sigma_{mr}$ is independent of velocity and is the approximate
cross section in the moderately relativistic limit (where the second
factor goes to $v$). Here $\delta$ is the splitting of the excited
state from the ground state and $M_\chi$ is the mass of the wimp.  The
threshold velocity for excitation is
$v_{thresh}=\sqrt{4\delta/M_\chi}$, so the integral becomes
\begin{eqnarray}
\sigmav(r) & = 8\pi^2\sigma_{mr} \int_0^\infty v^2f(v,r)dv 
   \int_0^\infty v'^2f(v',r)dv' \nonumber \\
   & \times \int_{-1}^1 d(\cos\theta) \sqrt{v_{rel}^2 - v_{thresh}^2}
\end{eqnarray}
with
\be
v_{rel}^2 = v^2 + v'^2 + 2vv'\cos\theta
\ee
This conveniently separates the problem into a question of particle
physics ($\sigma_{mr}$) and a question of astrophysics (the
phase-space densities).

\subsection{DM velocity distribution}
The WIMP velocity distribution depends on the details of the formation
history of the Galaxy, but for simplicity we use a Boltzmann
approximation.  Although there may be some WIMPs in the inner Galaxy
moving faster than escape velocity, it is reasonable to assume that
the distribution goes to zero rapidly for $v > v_{esc}(r)$. 
Therefore, we adopt
\hskip 0.3in
\be
f(v,r)=
\begin{cases}
N\exp(-v^2/2v_{rms}^2) & v \le v_{esc} \\
0                      & v > v_{esc}
\end{cases}
\ee
with 1-D velocity dispersion $v_{rms} = 200\kms$.  $N$ normalizes
this so that
$\int^{v_{esc}} d^3v f(v,r) =1$.
Following Merritt et al. \cite{Merritt:2005} we use a DM halo profile
of the form
\bel{eqn:merritt}
\rho = \rho_0 \exp\left[-\frac{2}{\alpha}
  \left(\frac{r^\alpha - R_\odot^\alpha}{r_{-2}^\alpha}\right)\right]
\ee
where $\rho_0 = 0.1-0.7~\gev$ \cite{PDBook} is the DM density at the Solar circle
($r=R_\odot$), $r_{-2}=25\kpc$ is the radius at which the logarithmic
slope of the profile is -2, and $\alpha \approx 0.2$ is a parameter of
the profile.  This profile is inspired by a fitting formula for the
logarithmic slope of density\cite{Navarro:2004}:
\be
\frac{d\ln\rho}{d\ln r} = -2(r/r_{-2})^\alpha
\ee
and agrees well with simulations \cite{Merritt:2005}.

This profile only contains the dark matter, which does not dominate in 
the inner parts of the Galaxy.
To estimate the escape velocity of the Milky Way, we note that the
rotation curve is nearly flat with a circular velocity $v_c\approx
220\kms$ out to far beyond the solar circle \cite{Fich:1989}.  In the Merritt profile the scale radius of $r_{-2}$
contains much of the mass of the Galaxy, so we make the (conservative)
approximation that the rotation curve is flat to $20\kpc$ and then there
is no mass beyond that radius.  A flat rotation curve implies that
enclosed mass $M(r)=r v_{c}^2/G$, so $\rho(r)\sim r^{-2}$.  This density
profile differs from the Merritt profile because it includes baryonic
matter, which dominates the mass density in the inner part of the Milky
Way.  (The Black Hole dominates only inside the inner 1 pc, so we
neglect its influence.)

The value of the one-dimensional velocity dispersion is chosen to agree with simulations \cite{Governato:2007}.  Although some authors have correctly pointed out
that in DM-only halos the dispersion decreases in the center
(e.g. \cite{Ascasibar:2006}), simulations including gas physics and
star formation indicate an increase \cite{Governato:2007}.  We defer a
detailed discussion of these and related issues to a future paper. 

The gravitational potential $\Phi = -v_c^2$ at $r=r_{-2}$, so
\be
\Phi(r) = -v_c^2-\int_r^{r_{-2}} \frac{GM(r')}{r'^2} dr'
= v_c^2[\ln(r/r_{-2})-1]
\ee
therefore, 
\be
v_{esc}(r) = \sqrt{-2\Phi(r)} = 
\begin{cases}
v_c\sqrt{2[1-\ln(r/r_{-2})]} & \text{for~} r<r_{-2} \\
v_c\sqrt{2r_{-2}/r}          & \text{for~} r > r_{-2}
\end{cases}
\ee
which for $v_c=220\kms$ gives $v_{esc}(8\kpc) = 431\kms$ and
$v_{esc}(0.5\kpc) = 674\kms$.  Thus an energy threshold for the
inelastic scattering corresponding to about $v_{rel}=1000\kms$ will cause a
sharp cutoff beyond a Galactocentric radius of 400 pc (about $3\degree$)
as observed by INTEGRAL.  Other assumptions may be made about
the mass profile outside of $20\kpc$, but since any additional
velocities add in quadrature, they do not modify this result
greatly.  
\newpage
\onecolumngrid

\begin{figure*}
\includegraphics[width=3in]{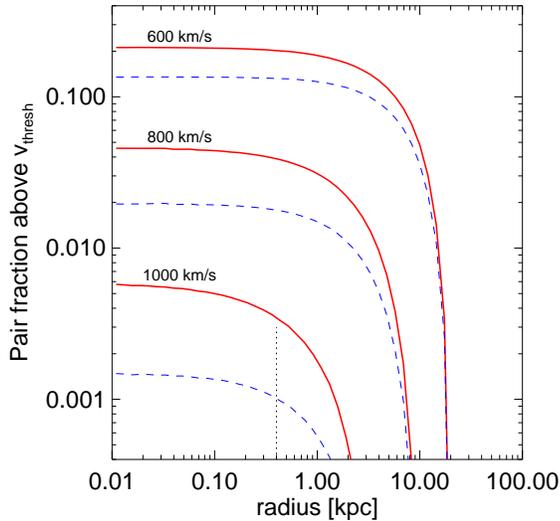}
\caption{\label{fig:velfrac}
The fraction of pairs with velocity above $v_{thresh}$ of 600, 800, or 1000 km/s, for $v_{rms}=200\kms$ (\emph{solid}) and $v_{rms}=180\kms$ (\emph{dashed}).  The approximate half-max radius of the observed 511 keV emission (\emph{vertical dotted}) is shown for reference. }
\end{figure*}
\twocolumngrid

\onecolumngrid

\begin{figure}
\includegraphics[width=6in]{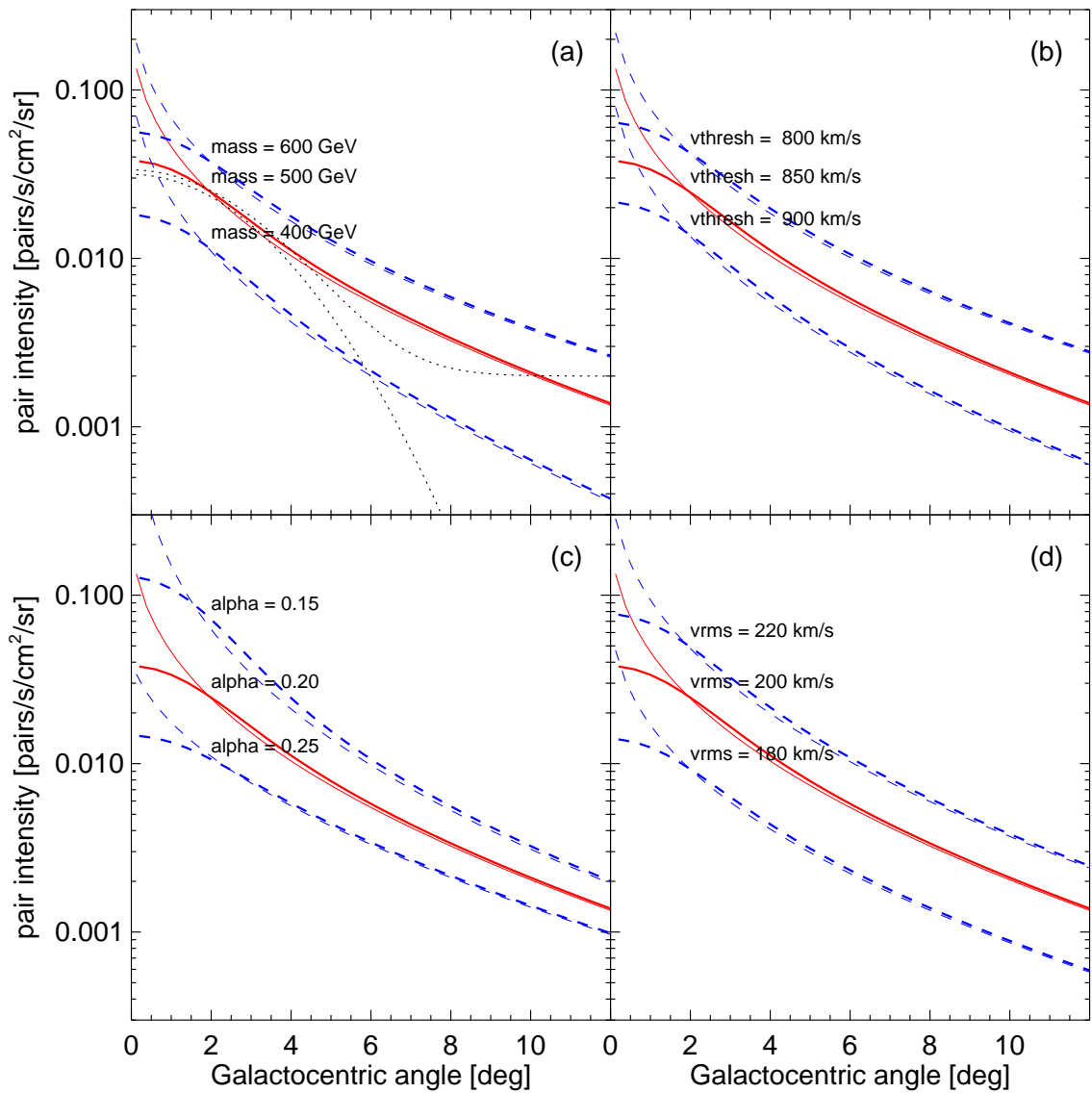}
\caption{\label{fig:spiflux4const}
Model curves for the constant sigma model in Eq (\ref{eqn:sigmavsimple}),
  taking $\sigma_{mr}=1.3\times 10^{-26}\cm^2\left(\frac{0.3~\gev/cm^3}{\rho_0}\right)^2= 2.4 \times 10^{-27} - 1.2 \times 10^{-25} \cm^2$, both unsmoothed (\emph{thin}) and 
  smoothed by a $3\degree$ FWHM beam (\emph{thick})
  to approximate the spatial response of SPI.  In all cases the solid
  red lines represent our fiducial model ($M=500~\gev$,
  $v_{thresh}=850\kms$, and halo parameters $v_{rms}=200\kms$, and Merritt index
  $\alpha=0.2$), and dashed blue lines represent variations of one
  parameter.
(a) $M=400, 500, 600~\gev$ with $\delta$ held fixed at $1~\Mev$ so that $v_{thresh}=950, 850, 776\kms$, respectively. 
(b) $v_{thresh}=800, 850, 900\kms$, keeping $M=500~\gev$ while $\delta$ now varies above and below $1~\Mev$. 
(c) The Merritt profile index $\alpha$ is varied (see Eq \ref{eqn:merritt}).
(d) The RMS velocity in the inner Milky Way is varied.  This is assumed to be constant with radius. 
  The observed $6\degree$ FWHM signal (\emph{lower dotted line}) and
  that signal plus an arbitrary baseline of 0.002 (\emph{upper dotted
  line}) are shown in (a).
The SPI zero is set by measurements $10\degree$ off the Galactic
  plane to correct for instrumental background.  }
\end{figure}
\twocolumngrid

\onecolumngrid

\begin{figure}
\includegraphics[width=6in]{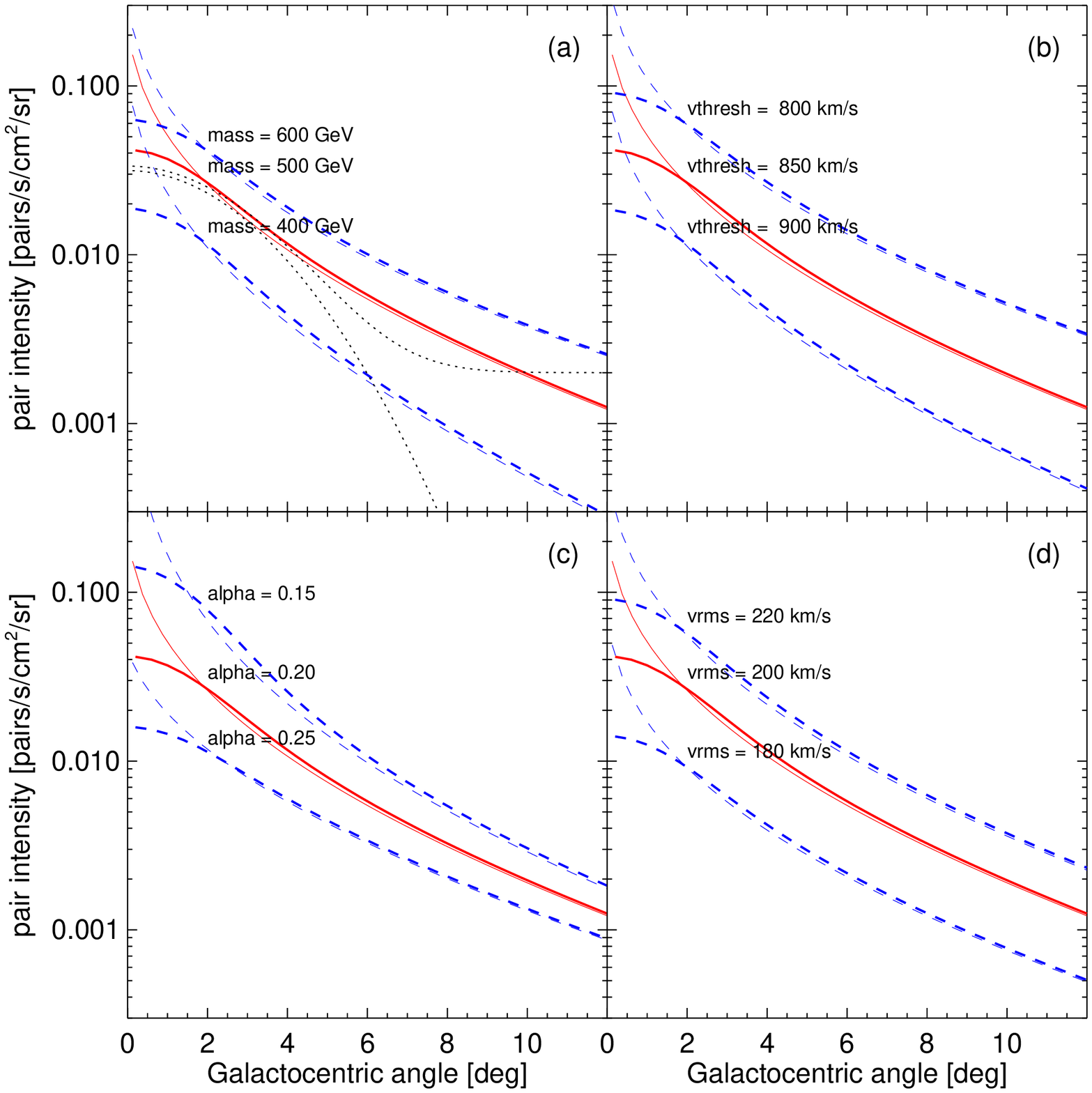}
\caption{\label{fig:spiflux4} 
Same as Fig. \ref{fig:spiflux4const},
  but using the full expression for the inelastic scattering cross
  section (Eq. \ref{eqn:kitchensink}) derived in Appendix A, setting the
  coupling $\lambda_+ = \lambda_- = 0.18$ and $\rho=0.4~\gev/{\rm cm}^3$.  }
\end{figure}

\twocolumngrid

It is instructive to consider the fraction of WIMP pairs with a
relative velocity above threshold, as a function of radius.  This
provides a sense of how many pairs are participating in the inelastic
scattering, independent of the detail of the particle physics model.
This pair fraction is shown in Fig. \ref{fig:velfrac} for three values
of $v_{thresh}$ and two values of $v_{rms}$, the 1-D velocity dispersion.

\subsection{Comparison with Observations}
To compare this model to the 511 keV map, we next integrate along the
line of sight
\be
\int_0^\infty dl \sigmav(r) n^2(r)
\ee
to obtain the number of pair creations per area per time.
Figs. \ref{fig:spiflux4const}, \ref{fig:spiflux4} show the line of sight integral for
various particle model and halo profile parameters.  For comparison, the most
recent analysis of the INTEGRAL SPI signal finds a $6\degree$ FWHM
Gaussian \cite{Weidenspointner:2007} in agreement with previous
measurements by OSSE \cite{Kinzer:2001ba}.
This corresponds to a pair intensity (inferred from both line and
Ps continuum) of
\be
I_{\rm pair} = \frac{L_{\rm pair}}{4\pi R_\odot^2} \frac{\exp(-\theta^2/2a^2)}{2\pi a^2}
\ee
where $L_{\rm pair} = 3\times 10^{42}$ pairs/s, $a=0.044$ rad (i.e.,
$6\degree /2.355$), $R_\odot=8\kpc$, and $\theta$ is the Galactocentric
angle.  This function peaks at $I\approx 0.03$
pairs$\s^{-1}\cm^{-2}\sr^{-1}$, and is shown in
Figs. \ref{fig:spiflux4const},\ref{fig:spiflux4} for comparison. In these plots, we have attempted to employ standard values from $\alpha,\rho_0$, and $v_{rms}$ for the fiducial model. However, by taking more optimal values for all parameters, a cross section of $\sigma_{mr} \simeq 3 \times 10^{-28}\cm^2$ would be sufficient to generate an acceptable rate to satisfy the INTEGRAL measurement.

None of the model lines in Fig. \ref{fig:spiflux4} are in perfect agreement with the
Gaussian, but keep in mind:
\begin{itemize}
\item The Gaussian form chosen by the SPI team is only an
      approximation.  A more thorough analysis would fit the SPI data
      directly to a   class of XDM models and compare likelihoods, as in
      \cite{Ascasibar:2006}. 

\item The substantial instrumental backgrounds in SPI are removed by
      subtracting observations of ``blank'' sky $10\degree$ off the
      Galactic plane.  We adjust the zero-point of the INTEGRAL
      measurements to match the model at $10\degree$, improving the agreement
      (see upper dotted line, Figs. \ref{fig:spiflux4const} and \ref{fig:spiflux4}).

\item We have chosen a narrow class of DM halo profiles for this
      benchmark model.  A different profile could improve
      agreement with the Gaussian.

\item Significant variation of the velocity dependence of the cross section could arise in different particle physics models

\item Other factors could alter the results, e.g. the \emph{elastic}
      scattering of WIMPs (see appendix) at some level may 
      alter the entropy in the
      inner part of the halo and cause significant departures from the
      assumed velocity and radial distributions.
\end{itemize}
Whatever deficiencies our model may suffer from in terms of the spatial
distribution of the positrons, they are less severe than annihilation
models (e.g. \cite{Boehm:2003bt}) because XDM provides a mechanism
for a radial cutoff.  Regardless of the details of $f(r,v)$ this is
an appealing aspect of XDM.

The question remains whether such a model of dark matter could naturally arise from particle physics considerations. In the following section, we will see that such models occur simply in extensions of the standard model.

\section{Models of Exciting Dark Matter}
\label{sec:model}
It is quite straightforward to find models of this sort. A model of composite dark matter, for instance, involving a bound state of constituent particles, would be expected to have precisely this sort of structure, arising from its excited states \footnote{Dark matter formed of a heavy doubly charged particle bound with a He nucleus would also have excitations of an appropriate scale \cite{Belotsky:2004ga,Fargion:2005xz,Fargion:2005ep,Khlopov:2005ew}. However, de-excitations would be via photons, not $e^+e^-$ pairs, so would not be a viable model of XDM.}. However, one needn't invoke complicated dynamics in order to generate a splitting, as an approximate symmetry will suffice.

An example of this is the neutron-proton mass difference, which is protected by an approximate isospin symmetry. Consequently, although the overall mass scale for the nucleons is GeV, the splitting is small, and protected against radiative corrections by the symmetry.
We will take a similar approach here.

Let us consider a Dirac fermion, composed of two Weyl fermions $\chi_{1,2}$. We will require the presence of a scalar field $\phi$, which will be light, and mediate the process of excitation. The Lagrangian we consider is
\bea
{\cal L}& =& \frac{1}{2} \partial_\mu \phi \partial^\mu \phi+\chi_{i}^\dagger \sigma_\mu \partial^\mu \chi_i - m_D \chi_1 \chi_2 \\&& \nonumber -\lambda_1 \phi  \chi_1 \chi_1 - \lambda_2 \phi \chi_2 \chi_2-V(\phi).
\eea
This can be justified with a ${\bf Z}_4$ symmetry, under which $\chi_{1,2} \rightarrow e^{\pm i \pi/2} \chi_{1,2}$ and $\phi \rightarrow - \phi$.

It is easiest to work in the basis $\chi,\chi_{*} = 1/\sqrt{2}(\chi_1 \mp \chi_2)$. In this basis the mass matrix for the $\chi$'s is
\be
M=
\begin{pmatrix}
\lambda_+ \phi - m_D & \lambda_- \phi \cr
\lambda_- \phi & \lambda_+ \phi + m_D
\end{pmatrix},
\ee
where $\lambda_\pm = \frac{1}{2}(\lambda_1 \pm \lambda_2)$.  There are
simple conclusions to draw from this expression. At leading order, the
Dirac fermion is understood as two degenerate Majorana fermions. If
the ${\bf Z}_4$ symmetry is broken weakly to ${\bf Z}_2$, for instance
by a $\phi$ expectation value, we expect these states to be split by a
small amount $\delta=2 \lambda_+ \vev{\phi}$. It should be noted that
because this is a symmetry breaking parameter, it can be naturally
much smaller than the mass of the dark matter particle. We
shall consider a simple origin for the vev shortly, but for now will
take it as a free parameter.

\begin{figure}
\begin{center}
a)
\scalebox{0.35}{
\fcolorbox{white}{white}{
  \begin{picture}(345,256) (60,-44)
    \SetWidth{2.5}
    \SetColor{Black}
    \ArrowLine(75,166)(225,121)
    \ArrowLine(225,121)(390,166)
    \ArrowLine(225,31)(390,-14)
    \ArrowLine(75,1)(225,31)
    \DashLine(225,121)(225,31){10}
    \Text(60,196)[lb]{\Large{\Black{$\chi$}}}
    \Text(375,196)[lb]{\Large{\Black{$\chi^*$}}}
    \Text(75,-29)[lb]{\Large{\Black{$\chi$}}}
    \Text(375,-44)[lb]{\Large{\Black{$\chi$}}}
    \Text(240,76)[lb]{\Large{\Black{$\phi$}}}
  \end{picture}
}
}\vskip0.2in
b)
\scalebox{0.35}{
\fcolorbox{white}{white}{
  \begin{picture}(585,256) (60,-44)
    \SetWidth{2.5}
    \SetColor{Black}
    \ArrowLine(75,166)(225,121)
    \ArrowLine(75,1)(225,31)
    \DashLine(225,121)(225,31){10}
    \Text(60,196)[lb]{\Large{\Black{$\chi$}}}
    \Text(75,-29)[lb]{\Large{\Black{$\chi$}}}
    \ArrowLine(450,121)(615,166)
    \Line(225,121)(450,121)
    \Line(225,31)(450,31)
    \ArrowLine(450,31)(615,-14)
    \Text(600,-44)[lb]{\Large{\Black{$\chi$}}}
    \Text(615,166)[lb]{\Large{\Black{$\chi^*$}}}
    \DashLine(270,121)(270,31){10}
    \DashLine(315,121)(315,31){10}
    \DashLine(360,121)(360,31){10}
    \Text(375,76)[lb]{\Large{\Black{$...$}}}
    \DashLine(450,121)(450,31){10}
  \end{picture}
}
}
\end{center}
\caption{\label{fig:excite}Excitation diagrams for $\chi \chi \rightarrow \chi^* \chi$.}
\end{figure}
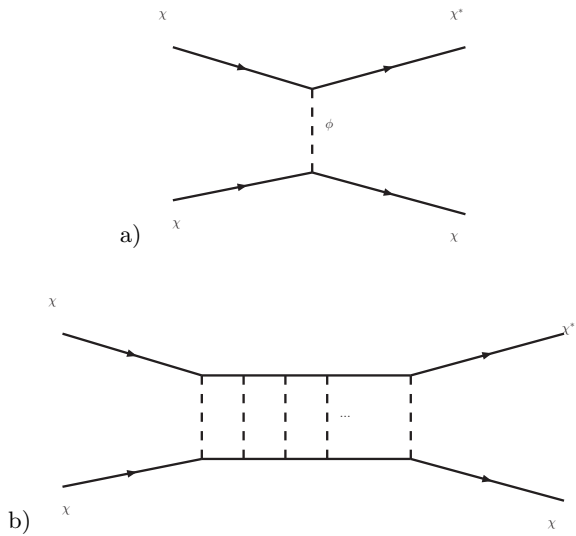
The process of excitation is mediated by the Feynman diagram of Fig.\ref{fig:excite}a. At very low velocity, one needs to sum the ladder diagrams \footnote{The authors strongly thank M. Pospelov for emphasizing this.}. The calculation of the cross section is given in the appendix. Qualitatively, however, the cross section is often approximately geometric, i.e., set by the scale of the characteristic momentum transfer. That is, $\sigma v \sim v/q^2 \sim v/m \delta \sim 10^{-19} \cm^3/\rm{s}$. This large excitation cross section, interestingly, does not occur for large coupling, but rather when the perturbative expansion in $\lambda^2/4 \pi^2 v$ breaks down.

The origin of the splitting will come from an expectation value of $\phi$. This can be derived from a potential of the form
\be
V(\phi) = -\frac{m_\phi^2}{2} \phi^2 + \frac{\kappa}{4} \phi^4
\ee
from which we yield a vacuum expectation value (vev) $\vev{\phi^2} = m_\phi^2/\kappa$. The mass splitting is then just $\delta = 2 \lambda_+ m_\phi/\sqrt{\kappa}$. 

The presence of such a vev seems to imply the existence of domain walls, which would be at the limit of what is allowed with regard to dominating the energy density of the universe \cite{kolbandturner}. This can be cured by the presence of cubic term $a \phi^3$, or the promotion of the symmetry to a (continuous) gauge symmetry. Although the cubic term is aesthetically unappealing, as our focus here is on the phenomenology of such a model with regard to present astrophysics, we shall accept such a term and defer more appealing solutions to future work.

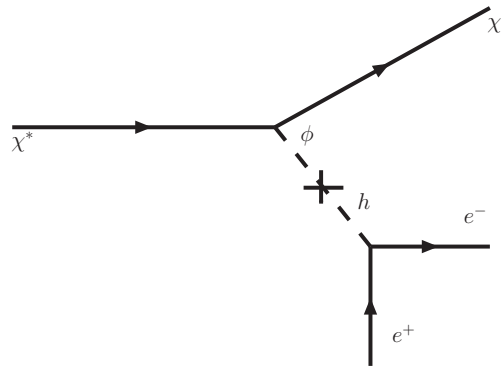
\begin{figure}
\begin{center}
\scalebox{0.6}{
\fcolorbox{white}{white}{
  \begin{picture}(330,226) (105,-44)
    \SetWidth{2.5}
    \SetColor{Black}
    \ArrowLine(105,106)(270,106)
    \ArrowLine(270,106)(405,181)
    \DashLine(270,106)(330,31){10}
    \ArrowLine(330,-44)(330,31)
    \ArrowLine(330,31)(405,31)
    \Text(105,91)[lb]{\Large{\Black{$\chi^*$}}}
    \Text(405,166)[lb]{\Large{\Black{$\chi$}}}
    \Text(345,-29)[lb]{\Large{\Black{$e^+$}}}
    \Text(390,46)[lb]{\Large{\Black{$e^-$}}}
    \Line(288,68)(313,68)
    \Line(299,79)(299,58)
    \Line(299,59)(299,60)
    \Text(287,95)[lb]{\Large{\Black{$\phi$}}}
    \Text(323,55)[lb]{\Large{\Black{$h$}}}
  \end{picture}
}
}
\end{center}
\caption{\label{fig:decay}Decay of the excited state into the ground state.}
\end{figure}

At this point we have only explained the process of excitation, but have not included the production of $e^+e^-$ pairs. This arises quite simply from a mixing term of the form 
\be
\alpha \phi^2 h h^*.
\ee
Such a term induces a mixing $\theta_{mix} \sim \alpha \vev{\phi}/m_h$. Thus, decays of the excited state into the lighter state proceed through the diagram in Fig. \ref{fig:decay}.

\subsection{XDM in the early universe}
While some worries, such as domain walls, can be cured simply, there are a number of non-trivial issues that must be addressed, most notably, the questions related to Big Bang Nucleosynthesis (BBN) and the relic abundance of the XDM.

Let us begin by addressing the lifetimes of the various particles. Both excited and unexcited states of the dark matter will come into thermal equilibrium in the early universe, with the excited state decaying into the lighter state with an approximate lifetime
\be
\tau_{\chi^*}^{-1} \sim \frac{ \lambda_-^2 \sin^2\theta_{mix} y_e^2 \delta^5}{\mu^4}
\ee
while the scalar $\phi$ decays with lifetime
\be
\tau_\phi^{-1} \approx \frac{m_\phi}{8 \pi} y_x^2 \sin^2\theta_{mix},
\ee
where $y_x$ is the Yukawa coupling of the heaviest fermion into which $\phi$ can decay. (In general, as $m_\phi \lsim 1 \gev$, this is $y_\mu$ or $y_e$.) With reasonable values for the mixing parameter, a scalar with mass $m\gsim 1 \mev$ will decay before nucleosynthesis. However, the decay of the excited state into the lighter state will naturally occur after nucleosynthesis, but well before recombination. Because of the low baryon to photon ratio, these positrons should thermalize before affecting the light element abundances, but it would be interesting to consider whether any effects would be observable.

Although structurally somewhat different, the relic abundance calculation for XDM is ultimately very similar to that of standard WIMPs. It is simplest, conceptually, if one separates the sectors of the theory into a dark sector ($\chi, \chi^*, \phi$) and the standard model sector. The relevant direct coupling between the two sectors is $\alpha$. Naturalness tells us that $\alpha \lsim m_\phi^{2}/v_{higgs}^2\lsim 10^{-4}$, so as not to significantly correct the $\phi$ vev. 

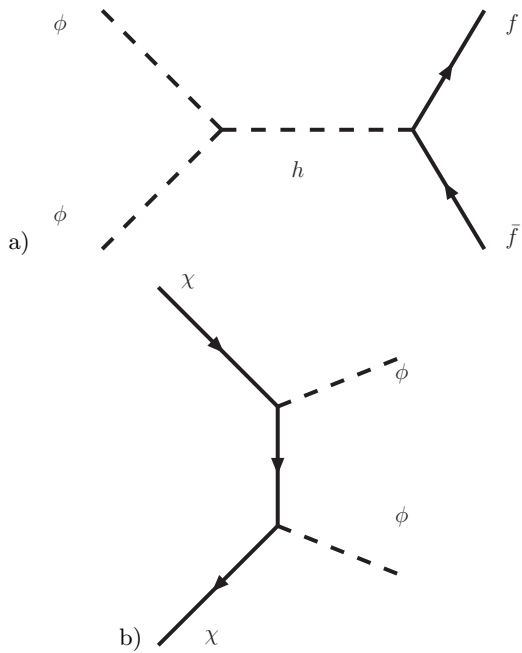
\begin{figure}
a)
\scalebox{0.6}{
\fcolorbox{white}{white}{
  \begin{picture}(315,151) (135,-59)
    \SetWidth{2.5}
    \SetColor{Black}
    \DashLine(165,91)(240,16){10}
    \DashLine(240,16)(360,16){10}
    \DashLine(240,16)(165,-59){10}
    \Text(135,76)[lb]{\Large{\Black{$\phi$}}}
    \Text(135,-44)[lb]{\Large{\Black{$\phi$}}}
    \Text(285,-14)[lb]{\Large{\Black{$h$}}}
    \ArrowLine(405,-59)(360,16)
    \ArrowLine(360,16)(405,91)
    \Text(420,-59)[lb]{\Large{\Black{$\bar f$}}}
    \Text(420,76)[lb]{\Large{\Black{$f$}}}
  \end{picture}
}
}\hskip 0.8in
b)\scalebox{0.6}{
\fcolorbox{white}{white}{
  \begin{picture}(180,241) (150,-74)
    \SetWidth{2.5}
    \SetColor{Black}
    \ArrowLine(150,151)(225,76)
    \ArrowLine(225,76)(225,1)
    \DashLine(225,76)(300,106){10}
    \DashLine(225,1)(300,-29){10}
    \ArrowLine(225,1)(150,-74)
    \Text(165,151)[lb]{\Large{\Black{$\chi$}}}
    \Text(180,-74)[lb]{\Large{\Black{$\chi$}}}
    \Text(300,91)[lb]{\Large{\Black{$\phi$}}}
    \Text(300,1)[lb]{\Large{\Black{$\phi$}}}
  \end{picture}
}
}
\caption{Diagrams contributing to thermal equilibrium in the dark sector and between the dark sector and the visible sector.}
\label{fig:ann}
\end{figure}

The relic abundance proceeds in a slight variation of usual WIMP freezeout. Here, $\chi$ will stay in equilibrium with $\phi$, which, in turn, stays in thermal equilibrium with the standard model. Direct scatterings $\phi \phi \leftrightarrow h h$ are in equilibrium at temperatures above the Higgs mass for $\alpha \gsim 10^{-8}$, thus we can reasonably tolerate $m_\phi \gsim 10 \mev$ from the naturalness constraint. In this way, the temperature of the dark sector is made equal to the temperature of the standard model particles. Whether this persists below the Higgs mass depends on $\alpha$. The s-channel annihilation into standard model fermions shown in Fig. \ref{fig:ann}a has a relativistic cross section $\sigma_{ann} = \alpha^2 m_f^2/(8 \pi m_h^4)$. This should keep annihilations in thermal equilibrium down to  $T=m_\mu, m_c \sim m_\tau,m_b$ for $\alpha \gsim 10^{-4},10^{-6}, 10^{-7}$, respectively.

 The annihilation rate of $\chi$ into $\phi$ occurs through the diagram in Fig \ref{fig:ann}b, with a cross section for $\chi \chi \rightarrow  \phi \phi$ annihilation of $\sigma_{ann} v_{rel} = 8 \lambda^4/( \pi M_\chi^2)$. The relic abundance from the freezeout of such an interaction is \cite{kolbandturner} (assuming equilibrium with the standard model) $\Omega h^2 \approx 2\times 10^{-11} {\rm GeV^{-2} }/\vev{\sigma_{ann} v_{rel}}$, which yields an acceptable relic abundance for $\lambda \sim \sqrt{ M/100 {\rm TeV}}$, or, roughly $\lambda \sim 1/10$, which is precisely the region in which the excitation process is maximal. This result should not be surprising - it is the usual result that a weak scale particle with perturbative coupling freezes out with the appropriate relic density. A rigorous study of the allowed parameter range is clearly warranted.

\subsection{Other decays and signals}
Although the freezeout process of XDM is similar to that of an
ordinary WIMP, the scattering and annihilation products can clearly be
different. One important feature is that the annihilation products are
typically electrons and neutrinos for $m_{\phi}<2 m_\pi$. However,
unlike, e.g., MeV dark matter, when the particles annihilate the
resulting electrons and positrons are extremely energetic. This is
intriguing because both the HEAT data \cite{Coutu:1999ws} and
the ``haze'' from the center of the galaxy \cite{Finkbeiner:2003im,
  Finkbeiner:2004us} point to new sources of multi-GeV electrons
and positrons.  Here, these high energy particles (from boosted
on-shell $\phi$ particles) are related to the low energy positrons detected
by INTEGRAL (from off-shell $\phi$ particles).  Such high energy particles may create high-energy gamma rays from inverse scattering off starlight which could be observed in the future GLAST mission. We leave detailed analysis of these possibilities for future work.

It should be noted that the electron and positron can be closed to allow a decay to two photons. Such a process is considerably suppressed, however, being down by a loop factor and suppressed by $\alpha$.  Consequently, these monochromatic photons should be produced at a rate roughly $10^{-5}$ that of the positron production rate, and are unlikely to be detectable within the diffuse background
(see, e.g., the discussion in \cite{Beacom:2005qv}).

\section{Discussion - other observable consequences}

\subsection{Cluster heating}
In the most massive galaxy clusters, the velocity dispersion is
high (up to $\sim 1000\kms$ 1-D), and the majority of pair velocities
may be above threshold.  It is doubtful
that the $e^+e^-$ signal could be seen by the current generation of
$\gamma-$ray telescopes, nor could the redshifted cosmological signal
be seen yet.  However, the kinetic energy of the pairs (which is
assumed to be $\la m_e$ when they are created) must
heat the intergalactic plasma.  In the intergalactic medium (IGM),
losses due to synchrotron
and inverse-Compton scattering are negligible at these energies.  Losses are 
dominated by bremsstrahlung and perhaps Alfven wave excitation (see
\cite{Loewenstein:1991}), and both of these mechanisms heat the few 
$\times 10^6$K plasma in the IGM.

The powerful X-ray emission from bremsstrahlung in galaxy clusters
(e.g. Perseus, Virgo) must either be balanced by significant
heating (perhaps from AGN) or else cooling matter must fall into the
cores.  Such cooling flows have been sought for years, but never found
at the level expected.  Recent X-ray spectroscopy has shown much less
emission at $\sim 1/3$ of the virial temperature than that expected
from simple radiative cooling models \cite{Peterson:2006}. 
Alternatives, such as light DM annihilation \cite{Ascasibar:2007}
have been suggested, but once again XDM can provide similar results. 
Especially noteworthy would
be models of XDM with $\sigmav$ rising as a higher power of $v$;
these could produce a substantial fraction of the power needed to
balance X-ray cooling in rich clusters.  Furthermore, in such
scenarios, the number of non-thermal (mildly relativistic) electrons
could distort the SZ effect in small but detectable ways.  Because
estimates of cluster heating and SZ signal are strongly
model-dependent, we defer discussion of this to a future paper.

\subsection{BH accretion}

The $z\sim6$ quasars discovered by the SDSS \cite{Fan:2003wd}
can be produced in simulations (Yuexing Li, priv. comm.) but only by
seeding the simulations with $200 M_\odot$ black holes at $z\sim30$.  A
mechanism of dissipative scattering, such as XDM,
provides the possibility of super-Eddington accretion.  
This could be important at high $z$, before baryons have had a
chance to cool into potential wells, perhaps seeding the universe with
black holes at early times.  At later times, however, the more
efficient cooling of baryons makes it unlikely that XDM accretion is a
significant factor in BH growth in general. 

\section{Conclusions}
We propose that dark matter is composed of WIMPs which can be
collisionally excited (XDM) and de-excite by $e^+e^-$ pair emission,
and that this mechanism is responsible for the majority of the
positronium annihilation signal observed in the inner Milky Way.  A
simple model of a pseudo-Dirac fermion with this property is presented,
though other models with similar phenomenology are likely
possible. The XDM framework is strongly motivated by a specific
problem, but has a rich set of implications for cluster heating, BH
formation, and other phenomena.  The model naturally freezes out with
approximately the right thermal relic density, and annihilations today
go mainly to $e^+e^-$.  Although XDM is dissipative,
it does not jeopardize stability of observed structure or disagree
with any other current observations.

\onecolumngrid
\appendix
\section{Excitation Cross Section}
The process of excitation arises from a sum of all ladder diagrams in addition to the tree level diagrams. We sum these using the spectator approach, where we take the intermediate fermion lines to be on-shell. Neglecting logarithmic dependencies on the momenta and scalar mass, one then finds the n$^{th}$ diagram to  be proportional to $q^{-2} ( \lambda^2/v)^n$, where $\lambda^2 = \lambda_{-,+}^2,\lambda_- \lambda_+$ depending on the diagram considered. Here $v$ is the velocity of {\em one} of the initial states $\chi$ in the center of mass frame. This allows us to sum the diagrams by solving
\be
M_{i,j \rightarrow k,l} =  i\frac{ \lambda_{i,k} \lambda_{j,l}}{q^2} + i \frac{L_{i,j,\alpha,\beta}  \lambda_{i,\alpha} \lambda_{j,\beta} M_{\alpha,\beta,k,l}}{v}
\ee
and equivalently for the u-channel (crossed) diagrams. Here $L$ is a loop factor which can depend on the states considered, and will be evaluated numerically. 

Because we are taking the intermediate fermion lines to be on-shell, we neglect diagrams where both fermions are excited, which are suppressed due to phase space.

\bel{eqn:kitchensink}
\sigma v_{rel} =\frac{2 v \beta  \left(v^2+\beta ^2\right)-\left(v^2-\beta
   ^2\right)^2 \tanh ^{-1}\left(\frac{2 v \beta
   }{v^2+\beta ^2}\right)}{4 m^2 \pi  v \left(v^2-\beta
   ^2\right)^2 \left(v^2+\beta ^2\right)}
   M^2
\ee
where $M^2$ arises from summing the internal single excitation ladder diagrams, excluding the external leg contributions to the amplitudes, which is
\bea
\nonumber M^2&=& 
\left(v^2 \lambda_-^2 \lambda _+^2 \left(\left(L_0-L_1\right)^2 \lambda _-^4-2 \lambda _+^2 \left(L_0-L_1\right)^2 \lambda_-^2+v^2+\lambda _+^4 \left(L_0-L_1\right)^2\right)\right)/ \\ &&
   \Big[ L_0^2 L_1^2 \lambda _+^8+2 \lambda _-^2 L_0^2 L_1^2 \lambda
   _+^6+\left(\left(L_1^2 \lambda _-^4+v^2\right) L_0^2+v^2 L_1^2\right) \lambda _+^4\\ &&+2 v^2 \lambda _-^2 \left(2 L_0-L_1\right) L_1
   \lambda _+^2+v^4+v^2 \lambda _-^4 L_1^2\Big] \nonumber
      \eea
where here, $L_{0,1}$ are the loop factors for diagrams involving zero or one excited internal fermions. $\beta$ is the final state velocity of one of the dark matter particles. To a reasonable approximation, $L_1$ is just suppressed relative to $L_0$ by phase space, that is, $L_1 \approx L_0 \beta^2/v^2$. With this simplification
\bea
&M^2 = v^2 \lambda _-^2 \lambda _+^2 \left(v^6+\left(v^2-\beta ^2\right)^2 \lambda _-^4 L_0^2+\left(v^2-\beta ^2\right)^2 \lambda _+^4
   L_0^2-2 \left(v^2-\beta ^2\right)^2 \lambda _-^2 \lambda _+^2 L_0^2\right)/& \\ 
   & \left(\beta ^4 L_0^2 \left(L_0^2 \lambda
   _+^4+v^2\right) \lambda _-^4+2 \beta ^2 \lambda _+^2 L_0^2 \left(2 v^4-\beta ^2 v^2+\beta ^2 \lambda _+^4 L_0^2\right) \lambda
   _-^2+\left(L_0^2 \lambda _+^4+v^2\right) \left(v^6+\beta ^4 \lambda _+^4 L_0^2\right)\right)& \nonumber
\eea
In these expressions, we have kept terms to leading order in $v,\beta$, and used the non-relativistic approximation.

We compute the factor $L_0$ numerically, and find $L_0 \approx 3 \times 10^{-2}$. Notice that in the perturbative limit, $M^2 = \lambda_-^2 \lambda_+^2$, as expected. This cross section peaks for $\lambda^2_{\pm}\approx v_{rel}/ L_0$ when the lowest order analysis begins to break down, which is $\lambda \sim 10^{-1}$. For $m= 500 \gev$, $\lambda_+=\lambda_-=0.18$ (the example used in the Fig. \ref{fig:spiflux4}) and $v = 450 \kms$ (i.e., $v_{rel} = 900 \kms$), one finds   $\sigma v_{rel} \sim .01 \gev^{-2}$. This is not a remarkable result, in that the exchanged $q^2$ is $O(0.5 \gev^2)$, so in this region, the cross section is approximately geometric. I.e., one would estimate a cross section $\pi/0.5^2$ - the particles will scatter if the come within approximately $d \sim 1/q$. It is important to note, however, that this large cross section does not occur in the region where the theory is strongly coupled (i.e., $\lambda_- \gg 1$), but rather in the region where one needs to sum the perturbative series, $\lambda^2/4 \pi v \sim 1$. Let us add that although the summation of the box diagrams does regulate the low-velocity properties of the scattering, when the cross section peaks (as a function of $\lambda$), every term in the summation is equally important, and it is likely that a more detailed analysis, including all the relevant logarithms is necessary.
\vskip 0.25in
\noindent {\bf Elastic Scattering Cross Section}
\vskip 0.1in
One can solve the equivalent equations to determine the elastic scattering cross section, as such a question may be relevant for questions of structure formation, as with the case of self-interacting dark matter. For the purposes of low-velocity elastic scattering, we can neglect excitation box diagrams, and one finds a total cross section
\be
\sigma \sim 1/m_\phi^2.
\ee
Given that $m_\phi$ can range from $1~\mev$ to $1~\gev$ we have cross sections from $10^{-28}\cm^2$ to $10^{-21}\cm^2$. With the very largest cross sections, the mean-free path of dark matter in the halo would be of the order of Mpc, which could be relevant for the questions of halo substructure \cite{2000PhRvL..84.3760S}. However, these scatterings are at very low momentum transfer, with $\Delta \theta \sim m_\phi/M$. Whether such small scatterings are significant enough to realize the scenario of Spergel and Steinhardt is worthy of additional study.

\vskip 0.35 in

\noindent {\bf Note Added:} 
After an earlier draft of this work was completed, we became aware of concurrent work by M. Pospelov and A. Ritz \cite{Pospelov:2007xh}, in which a similar scenario was considered where the dark matter excites into charged states which decay via electron emission. In this model, excitation would occur via short-distance processes. Although the authors come to a negative conclusion as to the viability of such a scenario from those in this paper, we believe this is due to the lack of long distance excitations, and different assumptions about the halo model.

\vskip 0.2 in
\noindent {\bf Acknowledgments}
\vskip 0.05in
\noindent It is a pleasure to acknowledge conversations with Greg Dobler, Gia Dvali, Andrei Gruzinov,  Dan Hooper, Marc Kamionkowski, Ramesh
Narayan, and Michele Redi.  Jose Cembranos, Jonathan Feng, Manoj Kaplinghat, and Arvind Rajaraman provided healthy skepticism.  Discussions
with Nikhil Padmanabhan were most enlightening.  We thank Maxim Pospelov for discussions and pointing out the need to sum the ladder diagrams. We thank Spencer Chang and Celine Boehm for comments on an earlier draft. A conversation with Rashid Sunyaev during his visit to CfA stimulated this
research.  DF is partially supported by NASA LTSA grant NAG5-12972. 
NW is supported by NSF CAREER grant PHY-0449818 and DOE OJI grant \# DE-FG02-06ER41417.
\bibliography{xdm}

\begin{thebibliography}{42}
\expandafter\ifx\csname natexlab\endcsname\relax\def\natexlab#1{#1}\fi
\expandafter\ifx\csname bibnamefont\endcsname\relax
  \def\bibnamefont#1{#1}\fi
\expandafter\ifx\csname bibfnamefont\endcsname\relax
  \def\bibfnamefont#1{#1}\fi
\expandafter\ifx\csname citenamefont\endcsname\relax
  \def\citenamefont#1{#1}\fi
\expandafter\ifx\csname url\endcsname\relax
  \def\url#1{\texttt{#1}}\fi
\expandafter\ifx\csname urlprefix\endcsname\relax\def\urlprefix{URL }\fi
\providecommand{\bibinfo}[2]{#2}
\providecommand{\eprint}[2][]{\url{#2}}

\bibitem[{\citenamefont{{Johnson} et~al.}(1972)\citenamefont{{Johnson},
  {Harnden}, and {Haymes}}}]{Johnson:1972}
\bibinfo{author}{\bibfnamefont{W.~N.} \bibnamefont{{Johnson}},
  \bibfnamefont{III}}, \bibinfo{author}{\bibfnamefont{F.~R.}
  \bibnamefont{{Harnden}}, \bibfnamefont{Jr.}}, \bibnamefont{and}
  \bibinfo{author}{\bibfnamefont{R.~C.} \bibnamefont{{Haymes}}},
  \bibinfo{journal}{\apjl} \textbf{\bibinfo{volume}{172}}, \bibinfo{pages}{L1+}
  (\bibinfo{year}{1972}).

\bibitem[{\citenamefont{{Leventhal}}(1973)}]{Leventhal:1973}
\bibinfo{author}{\bibfnamefont{M.}~\bibnamefont{{Leventhal}}},
  \bibinfo{journal}{\apjl} \textbf{\bibinfo{volume}{183}},
  \bibinfo{pages}{L147+} (\bibinfo{year}{1973}).

\bibitem[{\citenamefont{{Share} et~al.}(1988)\citenamefont{{Share}, {Kinzer},
  {Kurfess}, {Messina}, {Purcell}, {Chupp}, {Forrest}, and
  {Reppin}}}]{Share:1988}
\bibinfo{author}{\bibfnamefont{G.~H.} \bibnamefont{{Share}}},
  \bibinfo{author}{\bibfnamefont{R.~L.} \bibnamefont{{Kinzer}}},
  \bibinfo{author}{\bibfnamefont{J.~D.} \bibnamefont{{Kurfess}}},
  \bibinfo{author}{\bibfnamefont{D.~C.} \bibnamefont{{Messina}}},
  \bibinfo{author}{\bibfnamefont{W.~R.} \bibnamefont{{Purcell}}},
  \bibinfo{author}{\bibfnamefont{E.~L.} \bibnamefont{{Chupp}}},
  \bibinfo{author}{\bibfnamefont{D.~J.} \bibnamefont{{Forrest}}},
  \bibnamefont{and} \bibinfo{author}{\bibfnamefont{C.}~\bibnamefont{{Reppin}}},
  \bibinfo{journal}{\apj} \textbf{\bibinfo{volume}{326}}, \bibinfo{pages}{717}
  (\bibinfo{year}{1988}).

\bibitem[{\citenamefont{Kinzer et~al.}(2001)}]{Kinzer:2001ba}
\bibinfo{author}{\bibfnamefont{R.~L.} \bibnamefont{Kinzer}}
  \bibnamefont{et~al.}, \bibinfo{journal}{Astrophys. J.}
  \textbf{\bibinfo{volume}{559}}, \bibinfo{pages}{282} (\bibinfo{year}{2001}).

\bibitem[{\citenamefont{{Atti{\'e}} et~al.}(2003)\citenamefont{{Atti{\'e}},
  {Cordier}, {Gros}, {Laurent}, {Schanne}, {Tauzin}, {von Ballmoos}, {Bouchet},
  {Jean}, {Kn{\"o}dlseder} et~al.}}]{Attie:2003}
\bibinfo{author}{\bibfnamefont{D.}~\bibnamefont{{Atti{\'e}}}},
  \bibinfo{author}{\bibfnamefont{B.}~\bibnamefont{{Cordier}}},
  \bibinfo{author}{\bibfnamefont{M.}~\bibnamefont{{Gros}}},
  \bibinfo{author}{\bibfnamefont{P.}~\bibnamefont{{Laurent}}},
  \bibinfo{author}{\bibfnamefont{S.}~\bibnamefont{{Schanne}}},
  \bibinfo{author}{\bibfnamefont{G.}~\bibnamefont{{Tauzin}}},
  \bibinfo{author}{\bibfnamefont{P.}~\bibnamefont{{von Ballmoos}}},
  \bibinfo{author}{\bibfnamefont{L.}~\bibnamefont{{Bouchet}}},
  \bibinfo{author}{\bibfnamefont{P.}~\bibnamefont{{Jean}}},
  \bibinfo{author}{\bibfnamefont{J.}~\bibnamefont{{Kn{\"o}dlseder}}},
  \bibnamefont{et~al.}, \bibinfo{journal}{\aap} \textbf{\bibinfo{volume}{411}},
  \bibinfo{pages}{L71} (\bibinfo{year}{2003}), \eprint{astro-ph/0308504}.

\bibitem[{\citenamefont{{Vedrenne} et~al.}(2003)\citenamefont{{Vedrenne},
  {Roques}, {Sch{\"o}nfelder}, {Mandrou}, {Lichti}, {von Kienlin}, {Cordier},
  {Schanne}, {Kn{\"o}dlseder}, {Skinner} et~al.}}]{Vedrenne:2003}
\bibinfo{author}{\bibfnamefont{G.}~\bibnamefont{{Vedrenne}}},
  \bibinfo{author}{\bibfnamefont{J.-P.} \bibnamefont{{Roques}}},
  \bibinfo{author}{\bibfnamefont{V.}~\bibnamefont{{Sch{\"o}nfelder}}},
  \bibinfo{author}{\bibfnamefont{P.}~\bibnamefont{{Mandrou}}},
  \bibinfo{author}{\bibfnamefont{G.~G.} \bibnamefont{{Lichti}}},
  \bibinfo{author}{\bibfnamefont{A.}~\bibnamefont{{von Kienlin}}},
  \bibinfo{author}{\bibfnamefont{B.}~\bibnamefont{{Cordier}}},
  \bibinfo{author}{\bibfnamefont{S.}~\bibnamefont{{Schanne}}},
  \bibinfo{author}{\bibfnamefont{J.}~\bibnamefont{{Kn{\"o}dlseder}}},
  \bibinfo{author}{\bibfnamefont{G.}~\bibnamefont{{Skinner}}},
  \bibnamefont{et~al.}, \bibinfo{journal}{A \& A}
  \textbf{\bibinfo{volume}{411}}, \bibinfo{pages}{L63} (\bibinfo{year}{2003}).

\bibitem[{\citenamefont{Kn{\"o}dlseder et~al.}(2003)}]{Knodlseder:2003sv}
\bibinfo{author}{\bibfnamefont{J.}~\bibnamefont{Kn{\"o}dlseder}}
  \bibnamefont{et~al.}, \bibinfo{journal}{Astron. Astrophys.}
  \textbf{\bibinfo{volume}{411}}, \bibinfo{pages}{L457} (\bibinfo{year}{2003}),
  \eprint{astro-ph/0309442}.

\bibitem[{\citenamefont{Weidenspointner et~al.}(2006)}]{Weidenspointner:2006nu}
\bibinfo{author}{\bibfnamefont{G.}~\bibnamefont{Weidenspointner}}
  \bibnamefont{et~al.}, \bibinfo{journal}{\aap} \textbf{\bibinfo{volume}{450}},
  \bibinfo{pages}{1013} (\bibinfo{year}{2006}), \eprint{astro-ph/0601673}.

\bibitem[{\citenamefont{{Teegarden} et~al.}(2005)\citenamefont{{Teegarden},
  {Watanabe}, {Jean}, {Kn{\"o}dlseder}, {Lonjou}, {Roques}, {Skinner}, {von
  Ballmoos}, {Weidenspointner}, {Bazzano} et~al.}}]{Teegarden:2005}
\bibinfo{author}{\bibfnamefont{B.~J.} \bibnamefont{{Teegarden}}},
  \bibinfo{author}{\bibfnamefont{K.}~\bibnamefont{{Watanabe}}},
  \bibinfo{author}{\bibfnamefont{P.}~\bibnamefont{{Jean}}},
  \bibinfo{author}{\bibfnamefont{J.}~\bibnamefont{{Kn{\"o}dlseder}}},
  \bibinfo{author}{\bibfnamefont{V.}~\bibnamefont{{Lonjou}}},
  \bibinfo{author}{\bibfnamefont{J.~P.} \bibnamefont{{Roques}}},
  \bibinfo{author}{\bibfnamefont{G.~K.} \bibnamefont{{Skinner}}},
  \bibinfo{author}{\bibfnamefont{P.}~\bibnamefont{{von Ballmoos}}},
  \bibinfo{author}{\bibfnamefont{G.}~\bibnamefont{{Weidenspointner}}},
  \bibinfo{author}{\bibfnamefont{A.}~\bibnamefont{{Bazzano}}},
  \bibnamefont{et~al.}, \bibinfo{journal}{\apj} \textbf{\bibinfo{volume}{621}},
  \bibinfo{pages}{296} (\bibinfo{year}{2005}), \eprint{astro-ph/0410354}.

\bibitem[{\citenamefont{Churazov et~al.}(2005)\citenamefont{Churazov, Sunyaev,
  Sazonov, Revnivtsev, and Varshalovich}}]{Churazov:2004as}
\bibinfo{author}{\bibfnamefont{E.}~\bibnamefont{Churazov}},
  \bibinfo{author}{\bibfnamefont{R.}~\bibnamefont{Sunyaev}},
  \bibinfo{author}{\bibfnamefont{S.}~\bibnamefont{Sazonov}},
  \bibinfo{author}{\bibfnamefont{M.}~\bibnamefont{Revnivtsev}},
  \bibnamefont{and}
  \bibinfo{author}{\bibfnamefont{D.}~\bibnamefont{Varshalovich}},
  \bibinfo{journal}{Mon. Not. Roy. Astron. Soc.}
  \textbf{\bibinfo{volume}{357}}, \bibinfo{pages}{1377} (\bibinfo{year}{2005}),
  \eprint{astro-ph/0411351}.

\bibitem[{\citenamefont{Weidenspointner et~al.}(2007)}]{Weidenspointner:2007}
\bibinfo{author}{\bibfnamefont{G.}~\bibnamefont{Weidenspointner}}
  \bibnamefont{et~al.}, \bibinfo{journal}{arXiv:astro-ph/0702621v}
  (\bibinfo{year}{2007}), \eprint{astro-ph/0702621v1}.

\bibitem[{\citenamefont{Prantzos}(2006)}]{Prantzos:2005pz}
\bibinfo{author}{\bibfnamefont{N.}~\bibnamefont{Prantzos}},
  \bibinfo{journal}{Astron. Astrophys.} \textbf{\bibinfo{volume}{449}},
  \bibinfo{pages}{869} (\bibinfo{year}{2006}), \eprint{astro-ph/0511190}.

\bibitem[{\citenamefont{Kalemci et~al.}(2006)\citenamefont{Kalemci, Boggs,
  Milne, and Reynolds}}]{Kalemci:2006bz}
\bibinfo{author}{\bibfnamefont{E.}~\bibnamefont{Kalemci}},
  \bibinfo{author}{\bibfnamefont{S.~E.} \bibnamefont{Boggs}},
  \bibinfo{author}{\bibfnamefont{P.~A.} \bibnamefont{Milne}}, \bibnamefont{and}
  \bibinfo{author}{\bibfnamefont{S.~P.} \bibnamefont{Reynolds}},
  \bibinfo{journal}{Astrophys. J.} \textbf{\bibinfo{volume}{640}},
  \bibinfo{pages}{L55} (\bibinfo{year}{2006}), \eprint{astro-ph/0602233}.

\bibitem[{\citenamefont{{Pl{\"u}schke}
  et~al.}(2002)\citenamefont{{Pl{\"u}schke}, {Cervi{\~n}o}, {Diehl},
  {Kretschmer}, {Hartmann}, and {Kn{\"o}dlseder}}}]{Plueschke:2002}
\bibinfo{author}{\bibfnamefont{S.}~\bibnamefont{{Pl{\"u}schke}}},
  \bibinfo{author}{\bibfnamefont{M.}~\bibnamefont{{Cervi{\~n}o}}},
  \bibinfo{author}{\bibfnamefont{R.}~\bibnamefont{{Diehl}}},
  \bibinfo{author}{\bibfnamefont{K.}~\bibnamefont{{Kretschmer}}},
  \bibinfo{author}{\bibfnamefont{D.~H.} \bibnamefont{{Hartmann}}},
  \bibnamefont{and}
  \bibinfo{author}{\bibfnamefont{J.}~\bibnamefont{{Kn{\"o}dlseder}}},
  \bibinfo{journal}{New Astronomy Review} \textbf{\bibinfo{volume}{46}},
  \bibinfo{pages}{535} (\bibinfo{year}{2002}).

\bibitem[{\citenamefont{Cass{\'e} et~al.}(2004)\citenamefont{Cass{\'e},
  Cordier, Paul, and Schanne}}]{Casse:2003fh}
\bibinfo{author}{\bibfnamefont{M.}~\bibnamefont{Cass{\'e}}},
  \bibinfo{author}{\bibfnamefont{B.}~\bibnamefont{Cordier}},
  \bibinfo{author}{\bibfnamefont{J.}~\bibnamefont{Paul}}, \bibnamefont{and}
  \bibinfo{author}{\bibfnamefont{S.}~\bibnamefont{Schanne}},
  \bibinfo{journal}{Astrophys. J.} \textbf{\bibinfo{volume}{602}},
  \bibinfo{pages}{L17} (\bibinfo{year}{2004}), \eprint{astro-ph/0309824}.

\bibitem[{\citenamefont{{Parizot} et~al.}(2005)\citenamefont{{Parizot},
  {Cass{\'e}}, {Lehoucq}, and {Paul}}}]{Parizot:2005}
\bibinfo{author}{\bibfnamefont{E.}~\bibnamefont{{Parizot}}},
  \bibinfo{author}{\bibfnamefont{M.}~\bibnamefont{{Cass{\'e}}}},
  \bibinfo{author}{\bibfnamefont{R.}~\bibnamefont{{Lehoucq}}},
  \bibnamefont{and} \bibinfo{author}{\bibfnamefont{J.}~\bibnamefont{{Paul}}},
  \bibinfo{journal}{\aap} \textbf{\bibinfo{volume}{432}}, \bibinfo{pages}{889}
  (\bibinfo{year}{2005}), \eprint{astro-ph/0411656}.

\bibitem[{\citenamefont{{Guessoum} et~al.}(2006)\citenamefont{{Guessoum},
  {Jean}, and {Prantzos}}}]{Guessoum:2006}
\bibinfo{author}{\bibfnamefont{N.}~\bibnamefont{{Guessoum}}},
  \bibinfo{author}{\bibfnamefont{P.}~\bibnamefont{{Jean}}}, \bibnamefont{and}
  \bibinfo{author}{\bibfnamefont{N.}~\bibnamefont{{Prantzos}}},
  \bibinfo{journal}{\aap} \textbf{\bibinfo{volume}{457}}, \bibinfo{pages}{753}
  (\bibinfo{year}{2006}), \eprint{astro-ph/0607296}.

\bibitem[{\citenamefont{Bertone et~al.}(2005)\citenamefont{Bertone, Hooper, and
  Silk}}]{Bertone:2004pz}
\bibinfo{author}{\bibfnamefont{G.}~\bibnamefont{Bertone}},
  \bibinfo{author}{\bibfnamefont{D.}~\bibnamefont{Hooper}}, \bibnamefont{and}
  \bibinfo{author}{\bibfnamefont{J.}~\bibnamefont{Silk}},
  \bibinfo{journal}{Phys. Rept.} \textbf{\bibinfo{volume}{405}},
  \bibinfo{pages}{279} (\bibinfo{year}{2005}), \eprint{hep-ph/0404175}.

\bibitem[{\citenamefont{Boehm et~al.}(2004)\citenamefont{Boehm, Hooper, Silk,
  Cass{\'e}, and Paul}}]{Boehm:2003bt}
\bibinfo{author}{\bibfnamefont{C.}~\bibnamefont{Boehm}},
  \bibinfo{author}{\bibfnamefont{D.}~\bibnamefont{Hooper}},
  \bibinfo{author}{\bibfnamefont{J.}~\bibnamefont{Silk}},
  \bibinfo{author}{\bibfnamefont{M.}~\bibnamefont{Cass{\'e}}},
  \bibnamefont{and} \bibinfo{author}{\bibfnamefont{J.}~\bibnamefont{Paul}},
  \bibinfo{journal}{Phys. Rev. Lett.} \textbf{\bibinfo{volume}{92}},
  \bibinfo{pages}{101301} (\bibinfo{year}{2004}), \eprint{astro-ph/0309686}.

\bibitem[{\citenamefont{Jean et~al.}(2006)}]{Jean:2005af}
\bibinfo{author}{\bibfnamefont{P.}~\bibnamefont{Jean}} \bibnamefont{et~al.},
  \bibinfo{journal}{Astron. Astrophys.} \textbf{\bibinfo{volume}{445}},
  \bibinfo{pages}{579} (\bibinfo{year}{2006}), \eprint{astro-ph/0509298}.

\bibitem[{\citenamefont{Beacom et~al.}(2005)\citenamefont{Beacom, Bell, and
  Bertone}}]{Beacom:2004pe}
\bibinfo{author}{\bibfnamefont{J.~F.} \bibnamefont{Beacom}},
  \bibinfo{author}{\bibfnamefont{N.~F.} \bibnamefont{Bell}}, \bibnamefont{and}
  \bibinfo{author}{\bibfnamefont{G.}~\bibnamefont{Bertone}},
  \bibinfo{journal}{Phys. Rev. Lett.} \textbf{\bibinfo{volume}{94}},
  \bibinfo{pages}{171301} (\bibinfo{year}{2005}), \eprint{astro-ph/0409403}.

\bibitem[{\citenamefont{Beacom and Yuksel}(2006)}]{Beacom:2005qv}
\bibinfo{author}{\bibfnamefont{J.~F.} \bibnamefont{Beacom}} \bibnamefont{and}
  \bibinfo{author}{\bibfnamefont{H.}~\bibnamefont{Yuksel}},
  \bibinfo{journal}{Phys. Rev. Lett.} \textbf{\bibinfo{volume}{97}},
  \bibinfo{pages}{071102} (\bibinfo{year}{2006}), \eprint{astro-ph/0512411}.

\bibitem[{\citenamefont{{Merritt} et~al.}(2005)\citenamefont{{Merritt},
  {Navarro}, {Ludlow}, and {Jenkins}}}]{Merritt:2005}
\bibinfo{author}{\bibfnamefont{D.}~\bibnamefont{{Merritt}}},
  \bibinfo{author}{\bibfnamefont{J.~F.} \bibnamefont{{Navarro}}},
  \bibinfo{author}{\bibfnamefont{A.}~\bibnamefont{{Ludlow}}}, \bibnamefont{and}
  \bibinfo{author}{\bibfnamefont{A.}~\bibnamefont{{Jenkins}}},
  \bibinfo{journal}{\apjl} \textbf{\bibinfo{volume}{624}}, \bibinfo{pages}{L85}
  (\bibinfo{year}{2005}), \eprint{astro-ph/0502515}.

\bibitem[{\citenamefont{{Yao} et~al.}(2006)\citenamefont{{Yao}, {Amsler},
  {Asner}, {Barnett}, {Beringer}, {Burchat}, {Carone}, {Caso}, {Dahl},
  {D'Ambrosio} et~al.}}]{PDBook}
\bibinfo{author}{\bibfnamefont{W.-M.} \bibnamefont{{Yao}}},
  \bibinfo{author}{\bibfnamefont{C.}~\bibnamefont{{Amsler}}},
  \bibinfo{author}{\bibfnamefont{D.}~\bibnamefont{{Asner}}},
  \bibinfo{author}{\bibfnamefont{R.}~\bibnamefont{{Barnett}}},
  \bibinfo{author}{\bibfnamefont{J.}~\bibnamefont{{Beringer}}},
  \bibinfo{author}{\bibfnamefont{P.}~\bibnamefont{{Burchat}}},
  \bibinfo{author}{\bibfnamefont{C.}~\bibnamefont{{Carone}}},
  \bibinfo{author}{\bibfnamefont{C.}~\bibnamefont{{Caso}}},
  \bibinfo{author}{\bibfnamefont{O.}~\bibnamefont{{Dahl}}},
  \bibinfo{author}{\bibfnamefont{G.}~\bibnamefont{{D'Ambrosio}}},
  \bibnamefont{et~al.}, \bibinfo{journal}{{Journal of Physics G}}
  \textbf{\bibinfo{volume}{33}}, \bibinfo{pages}{1+} (\bibinfo{year}{2006}),
  \urlprefix\url{http://pdg.lbl.gov}.

\bibitem[{\citenamefont{{Navarro} et~al.}(2004)\citenamefont{{Navarro},
  {Hayashi}, {Power}, {Jenkins}, {Frenk}, {White}, {Springel}, {Stadel}, and
  {Quinn}}}]{Navarro:2004}
\bibinfo{author}{\bibfnamefont{J.~F.} \bibnamefont{{Navarro}}},
  \bibinfo{author}{\bibfnamefont{E.}~\bibnamefont{{Hayashi}}},
  \bibinfo{author}{\bibfnamefont{C.}~\bibnamefont{{Power}}},
  \bibinfo{author}{\bibfnamefont{A.~R.} \bibnamefont{{Jenkins}}},
  \bibinfo{author}{\bibfnamefont{C.~S.} \bibnamefont{{Frenk}}},
  \bibinfo{author}{\bibfnamefont{S.~D.~M.} \bibnamefont{{White}}},
  \bibinfo{author}{\bibfnamefont{V.}~\bibnamefont{{Springel}}},
  \bibinfo{author}{\bibfnamefont{J.}~\bibnamefont{{Stadel}}}, \bibnamefont{and}
  \bibinfo{author}{\bibfnamefont{T.~R.} \bibnamefont{{Quinn}}},
  \bibinfo{journal}{\mnras} \textbf{\bibinfo{volume}{349}},
  \bibinfo{pages}{1039} (\bibinfo{year}{2004}), \eprint{astro-ph/0311231}.

\bibitem[{\citenamefont{{Fich} et~al.}(1989)\citenamefont{{Fich}, {Blitz}, and
  {Stark}}}]{Fich:1989}
\bibinfo{author}{\bibfnamefont{M.}~\bibnamefont{{Fich}}},
  \bibinfo{author}{\bibfnamefont{L.}~\bibnamefont{{Blitz}}}, \bibnamefont{and}
  \bibinfo{author}{\bibfnamefont{A.~A.} \bibnamefont{{Stark}}},
  \bibinfo{journal}{\apj} \textbf{\bibinfo{volume}{342}}, \bibinfo{pages}{272}
  (\bibinfo{year}{1989}).

\bibitem[{\citenamefont{{Governato} et~al.}(2007)\citenamefont{{Governato},
  {Willman}, {Mayer}, {Brooks}, {Stinson}, {Valenzuela}, {Wadsley}, and
  {Quinn}}}]{Governato:2007}
\bibinfo{author}{\bibfnamefont{F.}~\bibnamefont{{Governato}}},
  \bibinfo{author}{\bibfnamefont{B.}~\bibnamefont{{Willman}}},
  \bibinfo{author}{\bibfnamefont{L.}~\bibnamefont{{Mayer}}},
  \bibinfo{author}{\bibfnamefont{A.}~\bibnamefont{{Brooks}}},
  \bibinfo{author}{\bibfnamefont{G.}~\bibnamefont{{Stinson}}},
  \bibinfo{author}{\bibfnamefont{O.}~\bibnamefont{{Valenzuela}}},
  \bibinfo{author}{\bibfnamefont{J.}~\bibnamefont{{Wadsley}}},
  \bibnamefont{and} \bibinfo{author}{\bibfnamefont{T.}~\bibnamefont{{Quinn}}},
  \bibinfo{journal}{\mnras} \textbf{\bibinfo{volume}{374}},
  \bibinfo{pages}{1479} (\bibinfo{year}{2007}).

\bibitem[{\citenamefont{{Ascasibar} et~al.}(2006)\citenamefont{{Ascasibar},
  {Jean}, {B{\oe}hm}, and {Kn{\"o}dlseder}}}]{Ascasibar:2006}
\bibinfo{author}{\bibfnamefont{Y.}~\bibnamefont{{Ascasibar}}},
  \bibinfo{author}{\bibfnamefont{P.}~\bibnamefont{{Jean}}},
  \bibinfo{author}{\bibfnamefont{C.}~\bibnamefont{{B{\oe}hm}}},
  \bibnamefont{and}
  \bibinfo{author}{\bibfnamefont{J.}~\bibnamefont{{Kn{\"o}dlseder}}},
  \bibinfo{journal}{Mon. Not. Roy. Astron. Soc.}
  \textbf{\bibinfo{volume}{368}}, \bibinfo{pages}{1695} (\bibinfo{year}{2006}),
  \eprint{astro-ph/0507142}.

\bibitem[{\citenamefont{{Kolb} and {Turner}}(1990)}]{kolbandturner}
\bibinfo{author}{\bibfnamefont{E.~W.} \bibnamefont{{Kolb}}} \bibnamefont{and}
  \bibinfo{author}{\bibfnamefont{M.~S.} \bibnamefont{{Turner}}},
  \emph{\bibinfo{title}{{The early universe}}} (\bibinfo{publisher}{Frontiers
  in Physics, Reading, MA: Addison-Wesley}, \bibinfo{year}{1990}).

\bibitem[{\citenamefont{Coutu et~al.}(1999)}]{Coutu:1999ws}
\bibinfo{author}{\bibfnamefont{S.}~\bibnamefont{Coutu}} \bibnamefont{et~al.},
  \bibinfo{journal}{Astropart. Phys.} \textbf{\bibinfo{volume}{11}},
  \bibinfo{pages}{429} (\bibinfo{year}{1999}), \eprint{astro-ph/9902162}.

\bibitem[{\citenamefont{Finkbeiner}(2004{\natexlab{a}})}]{Finkbeiner:2003im}
\bibinfo{author}{\bibfnamefont{D.~P.} \bibnamefont{Finkbeiner}},
  \bibinfo{journal}{Astrophys. J.} \textbf{\bibinfo{volume}{614}},
  \bibinfo{pages}{186} (\bibinfo{year}{2004}{\natexlab{a}}),
  \eprint{astro-ph/0311547}.

\bibitem[{\citenamefont{Finkbeiner}(2004{\natexlab{b}})}]{Finkbeiner:2004us}
\bibinfo{author}{\bibfnamefont{D.~P.} \bibnamefont{Finkbeiner}}
  (\bibinfo{year}{2004}{\natexlab{b}}), \eprint{astro-ph/0409027}.

\bibitem[{\citenamefont{{Loewenstein} et~al.}(1991)\citenamefont{{Loewenstein},
  {Zweibel}, and {Begelman}}}]{Loewenstein:1991}
\bibinfo{author}{\bibfnamefont{M.}~\bibnamefont{{Loewenstein}}},
  \bibinfo{author}{\bibfnamefont{E.~G.} \bibnamefont{{Zweibel}}},
  \bibnamefont{and} \bibinfo{author}{\bibfnamefont{M.~C.}
  \bibnamefont{{Begelman}}}, \bibinfo{journal}{\apj}
  \textbf{\bibinfo{volume}{377}}, \bibinfo{pages}{392} (\bibinfo{year}{1991}).

\bibitem[{\citenamefont{{Peterson} and {Fabian}}(2006)}]{Peterson:2006}
\bibinfo{author}{\bibfnamefont{J.~R.} \bibnamefont{{Peterson}}}
  \bibnamefont{and} \bibinfo{author}{\bibfnamefont{A.~C.}
  \bibnamefont{{Fabian}}}, \bibinfo{journal}{Phys. Reports}
  \textbf{\bibinfo{volume}{427}}, \bibinfo{pages}{1} (\bibinfo{year}{2006}),
  \eprint{astro-ph/0512549}.

\bibitem[{\citenamefont{{Ascasibar}}(2007)}]{Ascasibar:2007}
\bibinfo{author}{\bibfnamefont{Y.}~\bibnamefont{{Ascasibar}}},
  \bibinfo{journal}{\aap} \textbf{\bibinfo{volume}{462}}, \bibinfo{pages}{L65}
  (\bibinfo{year}{2007}), \eprint{astro-ph/0612130}.

\bibitem[{\citenamefont{Fan et~al.}(2003)}]{Fan:2003wd}
\bibinfo{author}{\bibfnamefont{X.}~\bibnamefont{Fan}} \bibnamefont{et~al.}
  (\bibinfo{collaboration}{SDSS}), \bibinfo{journal}{Astron. J.}
  \textbf{\bibinfo{volume}{125}}, \bibinfo{pages}{1649} (\bibinfo{year}{2003}),
  \eprint{astro-ph/0301135}.

\bibitem[{\citenamefont{{Spergel} and
  {Steinhardt}}(2000)}]{2000PhRvL..84.3760S}
\bibinfo{author}{\bibfnamefont{D.~N.} \bibnamefont{{Spergel}}}
  \bibnamefont{and} \bibinfo{author}{\bibfnamefont{P.~J.}
  \bibnamefont{{Steinhardt}}}, \bibinfo{journal}{Physical Review Letters}
  \textbf{\bibinfo{volume}{84}}, \bibinfo{pages}{3760} (\bibinfo{year}{2000}),
  \eprint{astro-ph/9909386}.

\bibitem[{\citenamefont{Pospelov and Ritz}(2007)}]{Pospelov:2007xh}
\bibinfo{author}{\bibfnamefont{M.}~\bibnamefont{Pospelov}} \bibnamefont{and}
  \bibinfo{author}{\bibfnamefont{A.}~\bibnamefont{Ritz}}
  (\bibinfo{year}{2007}), \eprint{hep-ph/0703128}.

\bibitem[{\citenamefont{Belotsky et~al.}(2004)}]{Belotsky:2004ga}
\bibinfo{author}{\bibfnamefont{K.}~\bibnamefont{Belotsky}} \bibnamefont{et~al.}
  (\bibinfo{year}{2004}), \eprint{hep-ph/0411271}.

\bibitem[{\citenamefont{Fargion and Khlopov}(2005)}]{Fargion:2005xz}
\bibinfo{author}{\bibfnamefont{D.}~\bibnamefont{Fargion}} \bibnamefont{and}
  \bibinfo{author}{\bibfnamefont{M.}~\bibnamefont{Khlopov}}
  (\bibinfo{year}{2005}), \eprint{hep-ph/0507087}.

\bibitem[{\citenamefont{Fargion et~al.}(2006)\citenamefont{Fargion, Khlopov,
  and Stephan}}]{Fargion:2005ep}
\bibinfo{author}{\bibfnamefont{D.}~\bibnamefont{Fargion}},
  \bibinfo{author}{\bibfnamefont{M.}~\bibnamefont{Khlopov}}, \bibnamefont{and}
  \bibinfo{author}{\bibfnamefont{C.~A.} \bibnamefont{Stephan}},
  \bibinfo{journal}{Class. Quant. Grav.} \textbf{\bibinfo{volume}{23}},
  \bibinfo{pages}{7305} (\bibinfo{year}{2006}), \eprint{astro-ph/0511789}.

\bibitem[{\citenamefont{Khlopov}(2006)}]{Khlopov:2005ew}
\bibinfo{author}{\bibfnamefont{M.~Y.} \bibnamefont{Khlopov}},
  \bibinfo{journal}{Pisma Zh. Eksp. Teor. Fiz.} \textbf{\bibinfo{volume}{83}},
  \bibinfo{pages}{3} (\bibinfo{year}{2006}), \eprint{astro-ph/0511796}.

\end{thebibliography}
\bibliographystyle{apsrev}

\end{document}